\documentclass[pra,showkeys,twocolumn,
superscriptaddress,
 amsmath,
 amssymb,
 aps,
]{revtex4-2}

\usepackage[utf8]{inputenc}

\usepackage{graphicx}
\usepackage{dcolumn}
\usepackage{bm}
\usepackage{color}

\begin{document}

\title{Intermittent emission of particles from a Bose-Einstein condensate in a one-dimensional lattice}

\author{L. Q. Lai}
\email{lqlai@njupt.edu.cn}
\affiliation{School of Science, Nanjing University of Posts and Telecommunications, Nanjing 210023, China}

\author{Z. Li}
\email{leezhao@hnu.edu.cn}
\affiliation{School of Electronic Engineering, Chengdu Technological University, Chengdu 611730, China}

\author{Q. H. Liu}
\affiliation{School of Physics and Electronics, Hunan University, Changsha 410082, China}

\author{Y. B. Yu}
\affiliation{School of Physics and Electronics, Hunan University, Changsha 410082, China}

\date{\today}

\begin{abstract}

We investigate particle emission from a Bose-Einstein condensate with periodically modulated interactions in a one-dimensional lattice. Within perturbative analysis, which leads to instabilities for discrete modes, we obtain the main regimes where the system can emit a large particle jet, and find that the emission is distinctly intermittent rather than continuous. The time evolution of the trapped particles exhibits a stair-like decay, and a larger drive induces a more significant intermittency. We further shed light on the dynamics of the stimulating process, and demonstrate that instead of a real suspension, the intermittency represents a build-up stage of the system. The theoretical framework might be generalized to the explorations on multiple-site systems with analogous configurations and couplings, and offer new insights into other fundamental nonequilibrium problems.

\end{abstract}


\maketitle

\section{Introduction}

Cold atom experiments have enabled precise quantum coherent manipulations of interparticle interaction in many-body systems \cite{Bloch} and ingenious control of versatile unconventional configurations \cite{Chin,Eckardt}, which revealed a number of novel nonequlibrium quantum effects \cite{Polkovnikov1,Moon}. In recent years, matter-wave jet emission resembling fireworks \cite{Clark}, and the follow-ups \cite{Fu1,Zhang2,Feng,Fu2,Meznasic,Kim} have attracted intensive attention. Pairwise interactions were modulated by time-periodic drive in these seminal experiments, which induced exponentially amplified excitations in a Bose-Einstein condensate, and a large number of stimulated particles rapidly escaped from the trap, leading to a burst of jets along the radial directions.

There have been various theoretical works exploring several aspects of the highly nonequilibrium phenomena \cite{ Zhang2,Feng,Lellouch,Yan,Zhai,Chih,Lai1,Lai2}, from the dynamics of the stimulating process of the observed pair emission \cite{Yan,Zhai,Chih} and a unique single-particle emission \cite{Lai1}, to the characterizations of rapid density oscillations, typical threshold behavior \cite{Lai2} and high-harmonic generations \cite{Feng}. Within the infinite lattice \cite{Lai1,Lai2}, configurations with more sites in the trap would be particularly interesting, through which one can seed diverse initial fluctuations, and investigate the correlations between the particle emission and the leads. Accordingly, if one wanted to explore the competitions among different modes in the collective emission, extended models should be employed. From the theoretical point of view, one can configure synthetic traps, condensates and couplings based upon demands, to study the properties of the resulting particle jets.

Related intriguing topics, in some other contexts, on the dynamics of systems with more than one condensate have been extensively reported. Researchers focused on arrays of Bose-Einstein condensates \cite{Abdullaev,Polkovnikov2, Konotop, Tsukada,Dunningham}, two coupled condensates \cite{Chen, Riboli, Whitlock, Lee, Haroutyunyan, Pan, Salasnich, Gati, Topfer,Luo} and three coupled condensates \cite{Nemoto, Zhang1, Franzosi1, Franzosi2, Buonsante1,Buonsante2, Guo, Cao}, and revealed many appealing phenomena, such as resonances, phase fluctuations and interference effects. Geometries with more sites and leads would somewhat make the analysis more complicated. For simplicity, in the present work we focus on a one-dimensional infinite lattice, where the trap confines three sites, and extensively explore the collective particle emission from a Bose-Einstein condensate under periodic drive in a transparent way, which can be generalized to multiple-site systems sharing similar configurations and couplings. Unlike the commonly observed and widely studied continuous emission, we find the distinctly intermittent emission process, where the trapped particles exhibit a stair-like decay. In particular, we revisit the previous two-site model \cite{Lai2} to further clarify the intermittency.

The rest of the paper is organized as follows. In Sec.~\ref{sec:model} we introduce our model and the relevant equations of motion. In Sec.~\ref{sec:perturb} we outline the perturbative analysis with respect to different modes, and discuss the regimes of instabilities. In Sec.~\ref{sec:numerics} we parametrically drive the system and present the numerical results. A summary is given in Sec.~\ref{sec:summary}.

\section{Theoretical Model}\label{sec:model}

\begin{figure}[htbp]
\includegraphics[width=1.0\columnwidth]{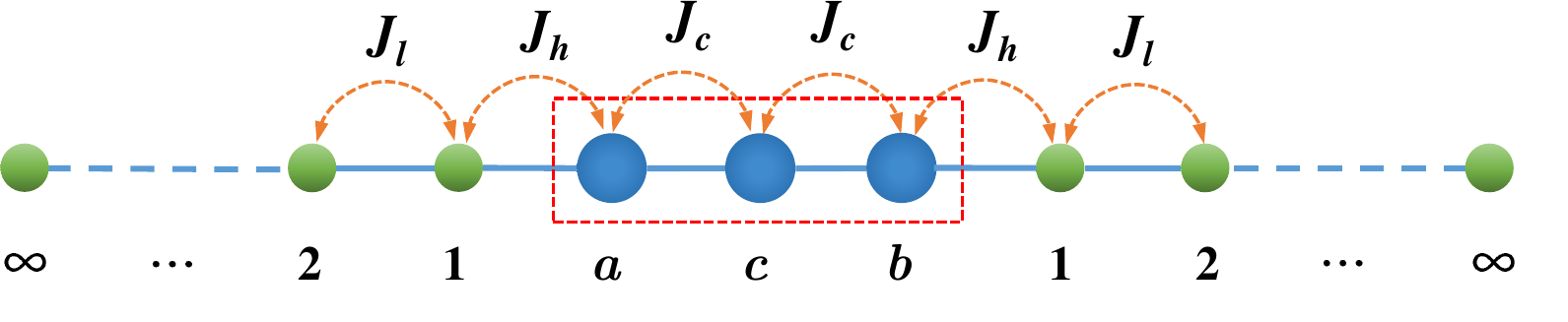}
\caption{Schematic of the infinite lattice. The red dashed box contains three locally trapped sites $a$, $b$ and $c$, and the sites on the leads are labeled by nonzero integers.}
\label{latticepic}
\end{figure}

As depicted in Fig.~\ref{latticepic}, we consider a one-dimensional infinite lattice, where the three central sites labeled $a$, $b$ and $c$ represent a confined Bose-Einstein condensate in a local deep trap of depth $V$. The coupling strength $J_{c}$ enables particles to hop back and forth among sites, while $J_h$ and $J_l$ quantify the hoppings from the trap to the leads and the couplings between nearest-neighbour sites in each lead, respectively. We assume that only interactions between atoms which sit on the central sites are included, and excited atoms with sufficient energy can move off to infinity along the leads, with lattice sites labeled by nonzero numbers $1,2,\ldots,\infty$. Such a configuration of symmetric geometry, with inhomogeneous lattice,  trap and localized interactions, could be literally implemented in an experiment involving optical lattices and microtraps \cite{Chin,Kuhr,Kaufman}.

This model can be described by the Hamiltonian
\begin{eqnarray}
\hat{H} &=& V\left( \hat{a}_{0}^{\dag }\hat{a}_{0}+\hat{b}_{0}^{\dag }\hat{b}_{0}+\hat{c}_{0}^{\dag }\hat{c}_{0}\right)  \nonumber \\
&&+\frac{1}{2}\left[ U+g\left(
t\right) \right] \left(
\hat{a}_{0}^{\dag }\hat{a}_{0}^{\dag }\hat{a}_{0}\hat{a}_{0}
+\hat{b}_{0}^{\dag }\hat{b}_{0}^{\dag }\hat{b}_{0}\hat{b}_{0}
+\hat{c}_{0}^{\dag }\hat{c}_{0}^{\dag }\hat{c}_{0}\hat{c}_{0}
\right) \nonumber \\
&&-J_{c}\left( \hat{a}_{0}^{\dag }\hat{c}_{0}+\hat{c}_{0}^{\dag }\hat{a}_{0}+\hat{b}_{0}^{\dag }\hat{c}_{0}+\hat{c}_{0}^{\dag }\hat{b}_{0}\right) \nonumber \\
&&-J_h \left(\hat{a}_1^\dag \hat{a}_0+\hat{a}_0^\dag \hat{a}_1+\hat{b}_1^\dag \hat{b}_0+\hat{b}_0^\dag \hat{b}_1 \right) \nonumber\\
&&-J_{l}\sum_{j=1}^{\infty }\left( \hat{a}_{j+1}^{\dag }\hat{a}_{j}+\hat{a}_{j}^{\dag }\hat{a}_{j+1}+\hat{b}_{j+1}^{\dag }\hat{b}_{j}+\hat{b}_{j}^{\dag }\hat{b}_{j+1}\right),
\end{eqnarray}
where $\hat{a}_{j}^{\dag}$ $(\hat{a}_{j})$ and $\hat{b}_{j}^{\dag}$ $(\hat{b}_{j})$ are creation (annihilation) operators on the $j$th site to the left and right, and $\hat{a}_0$, $\hat{b}_0$ and $\hat{c}_0$ represent the trapped sites. The term $U+g(t)$ characterizes the time-dependent pairwise interactions, where the on-site interaction $U$ is constant, and $g(t)=g\sin(\omega t)\theta(t)$ is the periodic drive, with $g$ the drive strength, $\omega$ the drive frequency, and $\theta(t)$ the step function.

By using the mean-field approach as discussed in Ref.~\cite{Lai1}, where the quantum and thermal fluctuations are neglected, we can thus write down the expectation value of the Heisenberg equations of motion for the central sites at $j=0$ ($\hbar=1$ throughout this paper)
\begin{eqnarray}
i\partial_{t}a_0 &=& \langle [\hat{a}_0,\hat{H}]\rangle  \nonumber \\
&=& Va_0+[U+g\sin(\omega t)]\vert a_0 \vert^2 a_0-J_{c} c_0-J_h a_1,\label{eoma0} \\
i\partial_{t}b_0 &=& \langle [\hat{b}_0,\hat{H}]\rangle \nonumber \\
&=& Vb_0+[U+g\sin(\omega t)]\vert b_0 \vert^2 b_0-J_{c} c_0-J_h b_1,\label{eomb0} \\
i\partial_{t}c_0 &=& \langle [\hat{c}_0,\hat{H}]\rangle \nonumber \\
&=& Vc_0+[U+g\sin(\omega t)]\vert c_0 \vert^2 c_0-J_{c}a_0-J_{c}b_0, \label{eomc0}
\end{eqnarray}
and for the sites where $j \geq 1$,
\begin{eqnarray}
i\partial_t a_1 &=& -J_h a_0-J_l a_2, \\
i\partial_t a_j &=& -J_l (a_{j+1}+a_{j-1}). \label{eomaj}
\end{eqnarray}
\begin{eqnarray}
i\partial_t b_1 &=& -J_h b_0-J_l b_2, \\
i\partial_t b_j &=& -J_l (b_{j+1}+b_{j-1}). \label{eombj}
\end{eqnarray}
In equilibrium where the driving field is absent, we begin from the ansatz $a_0=\alpha e^{-i \nu t}$, $b_0=\beta e^{-i \nu t}$, $c_0=\gamma e^{-i \nu t}$, where $\alpha$, $\beta$ and $\gamma$ are constant. Assuming that $a_1=\alpha e^{-i \nu t} e^{-\kappa_1}$ and $a_2=\alpha e^{-i \nu t} e^{-\kappa_1-\kappa}$. Eq.~(\ref{eomaj}) is a set of linear equations, i.e., $a_j/a_{j-1}=e^{-\kappa}$, thus $\cosh\kappa=\nu/(-2 J_l)$, and for $j\geq 1$ we have explicitly $a_j=\alpha e^{-i \nu t} e^{-\kappa_1} e^{-\kappa (j-1)}$  with $\kappa_{1}=-{\rm ln}(\frac{-J_h}{\nu+J_l e^{-\kappa}})$. Analogous equations for $b_{j\geq 1}$ are straightforward, while we only have $c_{j=0}$.

In the fireworks experiments \cite{Clark,Fu1,Fu2}, researchers generally maintained the dc component of the interparticle interaction strength as small, and in a previous work we have found that a finite $U$ did not qualitatively affect the related physics \cite{Lai2}, we thus take the limit where $U=0$, which leads to the nonlinear integro-differential equations,
\begin{eqnarray}
i\partial_t a_0&=&Va_0+g\sin(\omega t)\vert a_0 \vert^2 a_0-J_{c} c_0 \nonumber \\
&& +J_h^2\int^t G_{11}(t-\tau)a_0(\tau)d\tau,\label{mastera} \\
i\partial_t b_0&=&Vb_0+g\sin(\omega t)\vert b_0 \vert^2 b_0-J_{c} c_0 \nonumber \\
&& +J_h^2\int^t G_{11}(t-\tau)b_0(\tau)d\tau,\label{masterb} \\
i\partial_{t}c_0&=&Vc_0+g\sin(\omega t)\vert c_0 \vert^2 c_0-J_{c} (a_0+b_0), \label{masterc}
\end{eqnarray}
where $G_{j1}$ is the Green's function \cite{Lai1}
\begin{eqnarray}
G_{j1}(t)= i^{j-2} \frac{j \mathcal{J}_j(2 J_l t)}{J_l t} \theta(t)
\end{eqnarray}
with $\mathcal{J}_n(z)$ the Bessel function of the first kind. We will work at the perburbative level to largely simplify the analysis, where the drive strength $g$ and the coupling strength $J_h$ are both small.

\section{Perturbative Solutions}\label{sec:perturb}
We are now in position to perturbatively solve Eqs.~(\ref{mastera}) through (\ref{masterc}), and it is pretty straightforward to analytically analyze this model. We begin by finding the solutions with $g=0$, and in Appendix \ref{nonzero} we give the results for the case of a nonzero but small drive strength. With the ansatz in equilibrium, the equations become
\begin{eqnarray} \label{matrix}
\left(
\begin{array}{ccc}
\Delta_{\nu} & 0 & J_{c} \\
0 & \Delta_{\nu} & J_{c} \\
J_{c} & J_{c} & \nu -V
\end{array}
\right)
\left(
\begin{array}{c}
\alpha  \\
\beta  \\
\gamma
\end{array}
\right) =0
\end{eqnarray}
with $\Delta_{\nu}=\nu-V-J_{h}^{2}G_{11}(\nu)$. It is obvious that Eq.~(\ref{matrix}) has three eigen-solutions corresponding to three discrete modes in the system. From the symmetric case of $\alpha=\beta$, to the zeroth order of $J_{h}$ we directly get the ``M" mode with $\gamma_{\rm{M}}=-\sqrt{2}\alpha$ and $\nu_{\rm{M}}^{(0)}=V +\sqrt{2}J_c$, as well as the ``P" mode with $\gamma_{\rm{P}}=\sqrt{2}\alpha$ and $\nu_{\rm{P}}^{(0)}=V -\sqrt{2}J_c$. To the second order of $J_h$, simply substituting related $\gamma$ and $\nu$ into the equation leads to, respectively,
\begin{eqnarray} \label{nuM}
\nu_{\rm{M}}^{(2)} &=& V+\sqrt{2}J_c \nonumber \\
&& +\frac{J_h^2}{2J_l^2}\left(V+\sqrt{2}J_c-i\sqrt{4J_l^2-(V+\sqrt{2}J_c)^2}\right),
\end{eqnarray}
and
\begin{eqnarray} \label{nuP}
\nu_{\rm{P}}^{(2)} &=& V-\sqrt{2}J_c \nonumber \\
&& +\frac{J_h^2}{2J_l^2}\left(V-\sqrt{2}J_c-i\sqrt{4J_l^2-(V-\sqrt{2}J_c)^2}\right).
\end{eqnarray}
As for the antisymmetric case of $\alpha=-\beta$, similarly we can obtain the ``Z" mode with $\gamma_{\rm{Z}}=0$, $\nu_{\rm{Z}}^{(0)}=V$, and
\begin{eqnarray} \label{nuZ}
\nu_{\rm{Z}}^{(2)}&=&V+\frac{J_{h}^2}{2J_{l}^2}(V-i\sqrt{4J_{l}^2-V^2}).
\end{eqnarray}

To observe significant particle emission, we need to be in the regime where the ``P" mode and the ``Z" mode are stable but the ``M" mode is damped, i.e., $\vert V-\sqrt{2}J_c \vert>2J_l$, $\vert V \vert>2J_{l}$ and $\vert V+\sqrt{2}J_c \vert<2J_l$. Specializing to the case that $V=-\vert V \vert<0$, we reach the constraints of the allowed values for the trapping potential as $-\sqrt{2}J_{c}-2J_{l}<V<-2J_{l}$, under which the particles of a large-amplitude ``M" mode can decay into jets.

\section{Numerical Results}\label{sec:numerics}

Pair atoms excited out of the trap share half of the driving energy, and move along the leads in different directions, resulting in particle jets. In the following, we will parametrically drive the system by modulating the interaction strength and numerically solving Eqs.~(\ref{mastera}) through (\ref{masterc}). We assume that for $t<0$ the system is in equilibrium, while the perturbation is turned on at time $t=0$ with $g(t)=g\sin(\omega t)$. Without any loss of generality, we seed the modes by taking $\alpha \approx \beta=1$, with a slight difference in the calculations. From the related ``M" mode we take $\gamma=\sqrt{2}$, i.e., all the particles are initially trapped in the stable ``P" mode before the system meets the conditions of collective emission under weak drives and proper frequencies. Though we preserve the scale of energy unit $J_{c}$ in some discussions as appropriate, in the numerics it is fixed as $J_{c}=1$.

\begin{figure}[htbp]
\includegraphics[width=1.0\columnwidth]{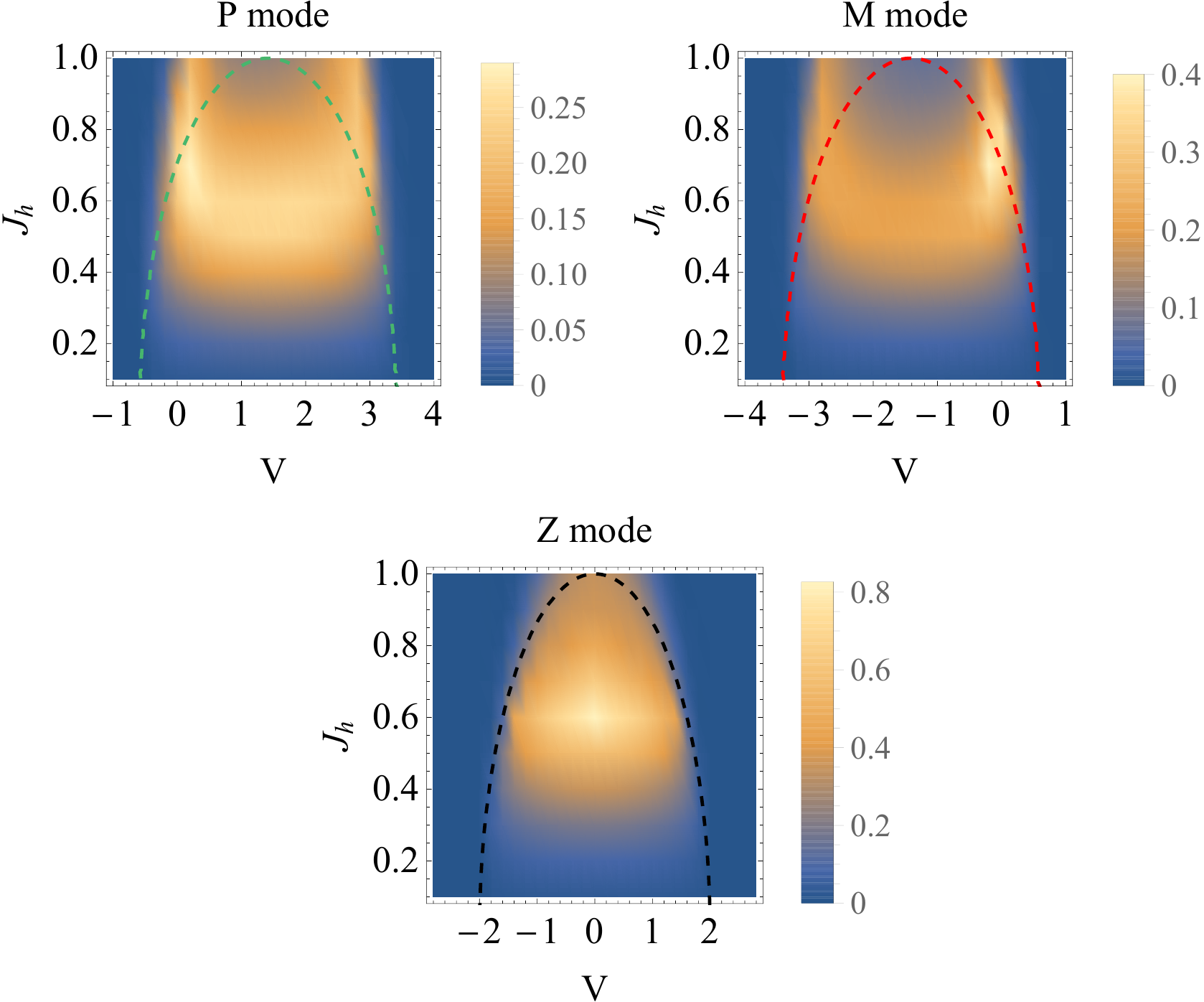}
\caption{(color online) Comparisons of the analytical (Dashed lines, contour plot) and numerical regimes (Orange areas, density plot) for the three modes. The analytical solutions are coming from Eqs.~(\ref{nuM}), (\ref{nuP}) and (\ref{nuZ}), and the numerical results are obtained by solving Eqs.~(\ref{mastera}) through (\ref{masterc}). Here, the drive strength is $g=0.1$, the simulation time is $t=25$, and the coupling strengths are $J_l=1$ and $J_c=1$. Color bars denote the average rate at which the particles are emitted from the trap. Energies are in units of $J_c$, and times are in units of $\hbar/J_c$.}
\label{modecom}
\end{figure}

The decay of the condensate is nonexponential, as the number of total particles in the central sites $N_{\rm{tot}}(t)=|a_{0}(t)|^2+|b_{0}(t)|^2+|c_{0}(t)|^2$ is a highly nonlinear function. We thus quantify the decay by fitting the total particle number to an exponential, $N_{\rm{tot}}=Ae^{-\Gamma t}$, where $\Gamma$ represents the average decay rate at which the atoms are ejected from the condensate. We first compare the analytical and numerical regimes for exciting the system, as shown in Fig.~{\ref{modecom}}, which clarifies how one could employ proper trapping potential $V$ and coupling strength $J_h$. For relatively small $J_h$, the instabilities of the three modes described by the numerics can be well delineated by the analytical solutions. When $J_h$ is larger (e.g., $J_h>0.5$), the deviations grow with the increasing $J_h$, under which circumstance the perturbative analysis might be inapplicable. For accuracy, we will mainly take $V=-3$ and $J_h=0.1$ hereafter, and also choose appropriate drive strength $g$ within the reasonable regimes.

\begin{figure}[htbp]
\includegraphics[width=1.0\columnwidth]{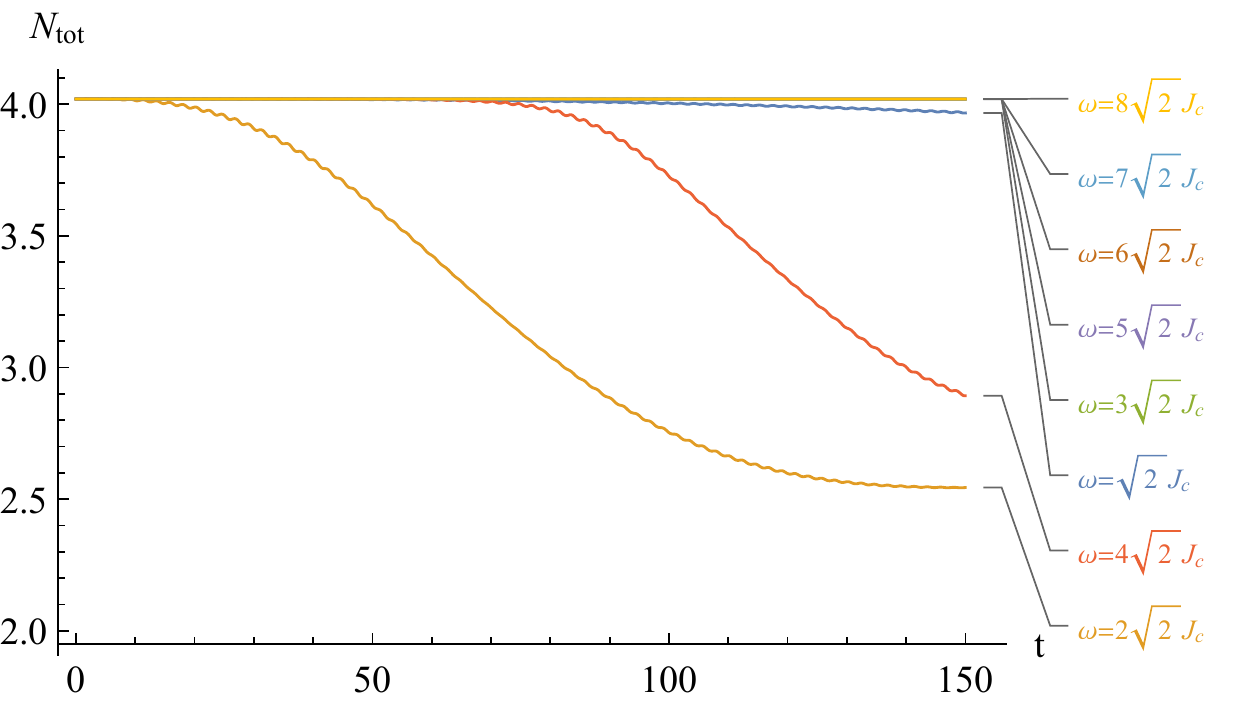}
\caption{(color online) Total particle number $N_{\rm{tot}}$ as a function of time under different drive frequencies $\omega$. Here, the trapping potential is $V=-3$, the drive strength is $g=0.1$, and the coupling strengths are $J_h=0.1$ and $J_l=1$. Energies are in units of $J_c$, and times are in units of $\hbar/J_c$.}
\label{frecom}
\end{figure}

Figure \ref{frecom} shows the short-time evolution of the trapped particles $N_{\rm{tot}}$ under typical frequencies $\omega$ in multiples of $\sqrt{2}J_{c}$, when the drive strength $g$ is fixed. As can be plainly seen, the system keeps fairly stable in the beginning for a short period of time, until the case with frequency $\omega=2\sqrt{2}J_{c}$ firstly decays significantly at about $t=25$. For the case with $\omega=4\sqrt{2}J_{c}$, at roughly $t=90$ the system also starts to decay rapidly, while emitting a somewhat less portion of particles than that of $\omega=2\sqrt{2}J_c$. Note that the characteristic is quite similar to that of the two-site model \cite{Lai2}: The trapped particles in the central sites are very stable unless the drive is resonant, but there is only one resonant frequency.

There are actually very narrow resonances in the spectrum if one specifically sweeps the frequency by tiny steps. The frequencies in between the resonances are also available for ejecting particles, while $\omega=2\sqrt{2}J_c$ and $\omega=4\sqrt{2}J_c$ correspond to distinct peaks that can lead to the most significant emission. Thus, we term them as the ``main frequencies". From the point on we will exclusively take $\omega=4\sqrt{2}J_{c}$ into account, which means that by choosing proper trapping potential the system is seeded at the lowest ``P" mode of $\nu_{\rm{P}}=V-\sqrt{2}J_c$, and under typical drive frequency $\omega=2[(V+\sqrt{2}J_c)-(V-\sqrt{2}J_c)]=4\sqrt{2}J_c$ it grows exponentially to the ``M" mode of $\nu_{\rm{M}}=V+\sqrt{2}J_c$, before decaying into significant jets. Once suitable driving conditions and a long enough simulation time were employed, the case under the other main frequency $\omega=2\sqrt{2}J_{c}$ would exhibit similar decays.

\begin{figure}[htbp]
\includegraphics[width=\columnwidth]{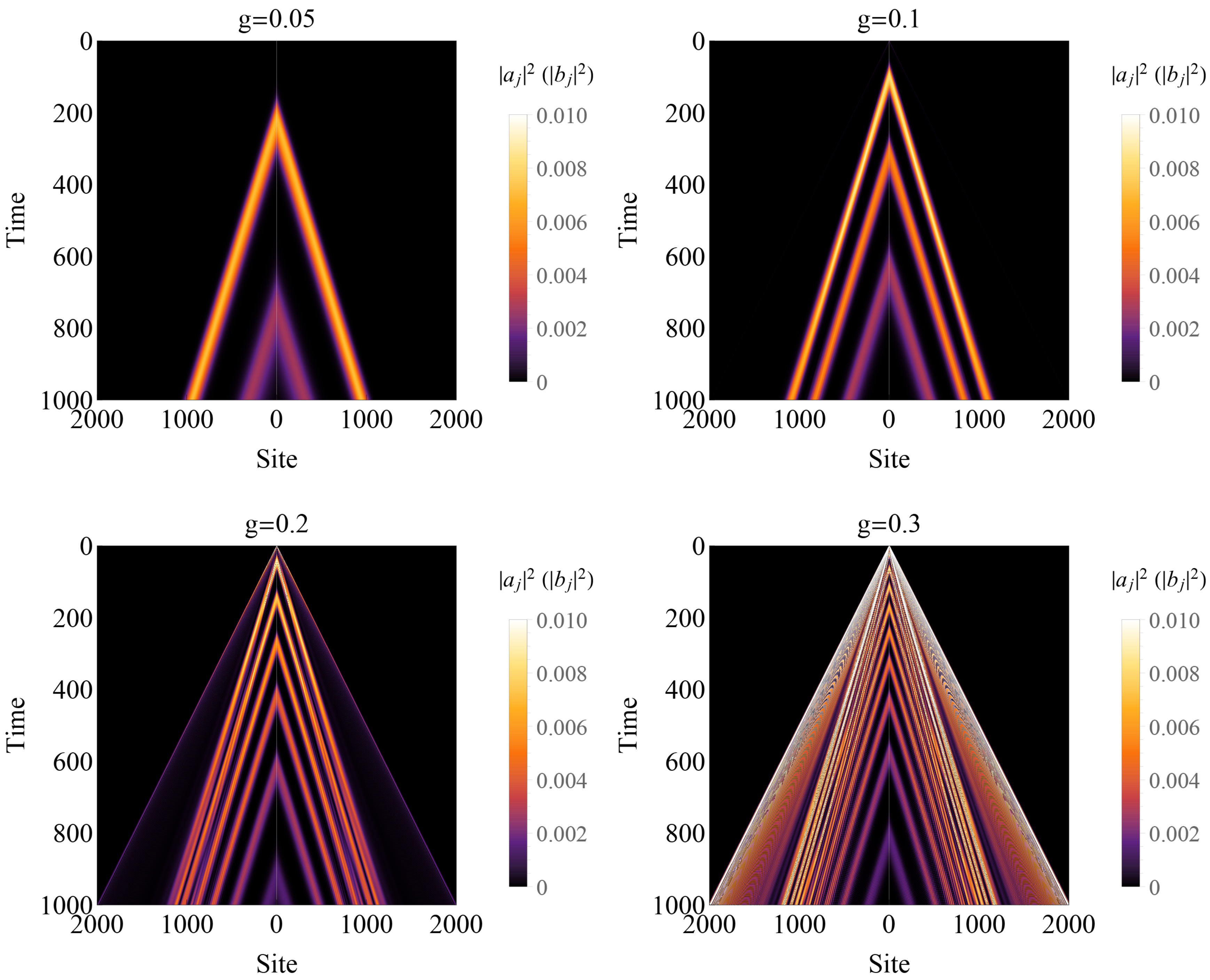}
\caption{(color online) Density distribution of particles on the $j$th site of each lead as a function of time. Here, the trapping potential is $V=-3$, the coupling strengths are $J_h=0.1$ and $J_l=1$, and the drive frequency is $\omega=4\sqrt{2}J_c$. Energies are in units of $J_c$, and times are in units of $\hbar/J_c$.}
\label{jetarraytot3grid}
\end{figure}

If we comprehensively consider the long-time dependence of particles on the sites of each lead, the density distributions will be straightforward, as demonstrated in Fig.~\ref{jetarraytot3grid}. For a relatively small drive strength ($g=0.05$), only a very small amount of particles are excited in the beginning, and most of them are still trapped. A continuous emission occurs at times $200<t<350$, with a large number of particles ejecting into the leads of the left and right. However, the emission seems to suddenly suspend with a ``pause" at about $t=400$, and lasts until $t=700$ when the emission restarts with a comparatively smaller decay rate. As for $g=0.1$, the response of the system is analogous to the case of $g=0.05$, while the emission emerges earlier, and it exhibits a first pause between $t=200$ and $t=300$, a second pause between $t=400$ and $t=600$, and a third pause at $t>750$, respectively. There are more pauses and more complicated situations for drives $g=0.2$ and $g=0.3$.

\begin{figure}[htbp]
\includegraphics[width=\columnwidth]{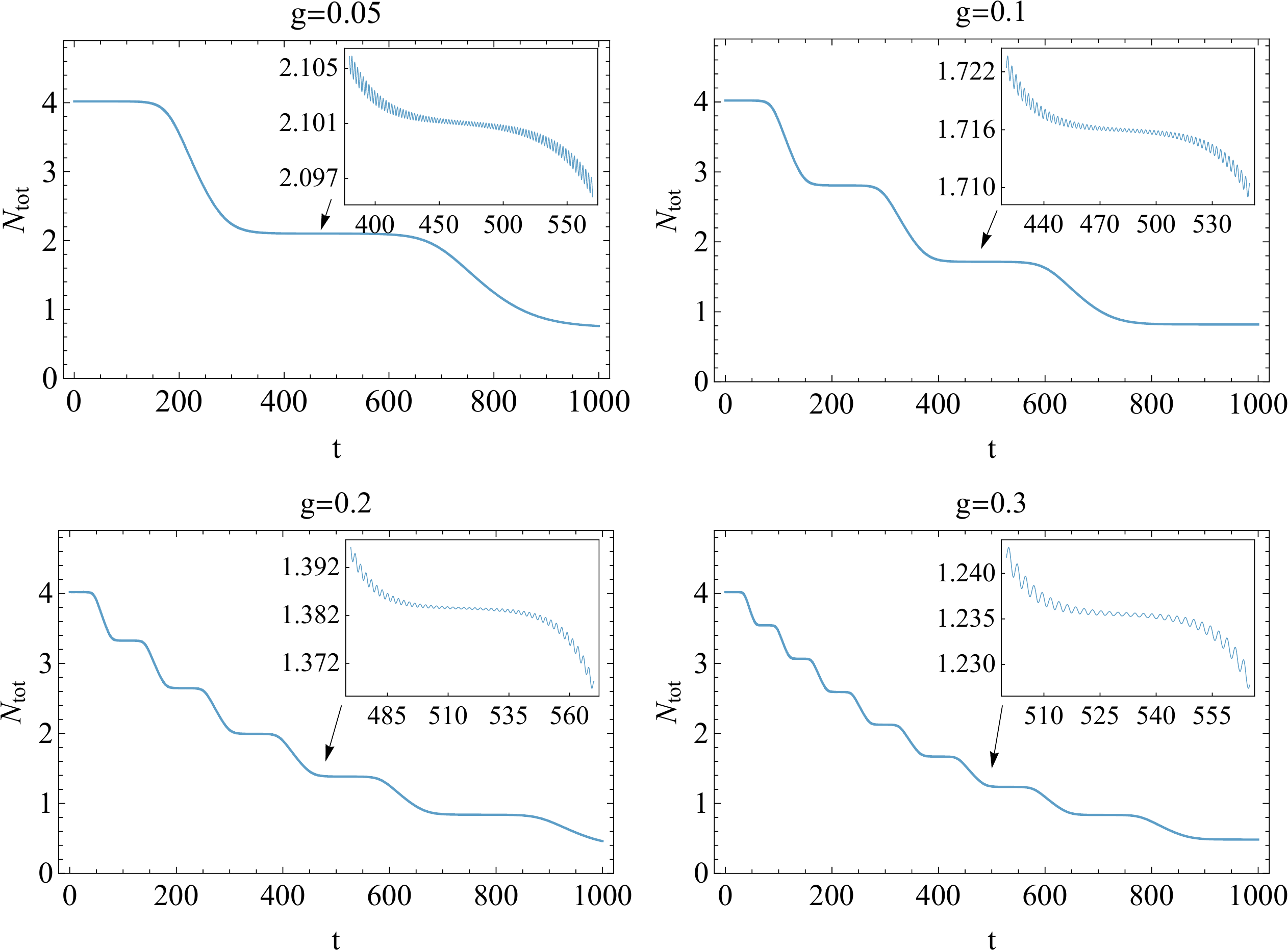}
\caption{(color online) The total particle number as a function of time under different drive strengths. The insets are the decaying details of the specified plateau. Here, the trapping potential is $V=-3$, the coupling strengths are $J_h=0.1$ and $J_l=1$, and the drive frequency is $\omega=4\sqrt{2}J_c$. Energies are in units of $J_c$, and times are in units of $\hbar/J_c$.}
\label{decaylong-insets-grid}
\end{figure}

For clarity, we present in Fig.~\ref{decaylong-insets-grid} more quantitatively the intermittency of the emission. One can clearly see that the decays of the trapped particles are stair-like. For drive strength $g=0.05$ the total particle number $N_{\rm tot}$ changes only a bit at $t<150$, while it drops dramatically afterwards, and a ``plateau" appears between $t=350$ and $t=550$, during which $N_{\rm tot}$ remains almost unchanged. Subsequently the system reemits particles with a relatively smaller rate. With respect to $g=0.1$, the $N_{\rm tot}$ also hardly changes in the beginning ($t<100$), until it decays successively to the second, the third and the fourth plateau, respectively. The third plateau has a right shift and less remaining particles compared with that of $g=0.05$ (the comparison of the insets). When even larger drives are exerted, i.e., $g=0.2$ and $g=0.3$, there would be more and more plateaus, where the times of appearance are staggered. Note that the duration of the first plateau of $g=0.3$ is the shortest among these drives, indicating an earlier emission under a larger drive.

\begin{figure}[htbp]
\includegraphics[width=\columnwidth]{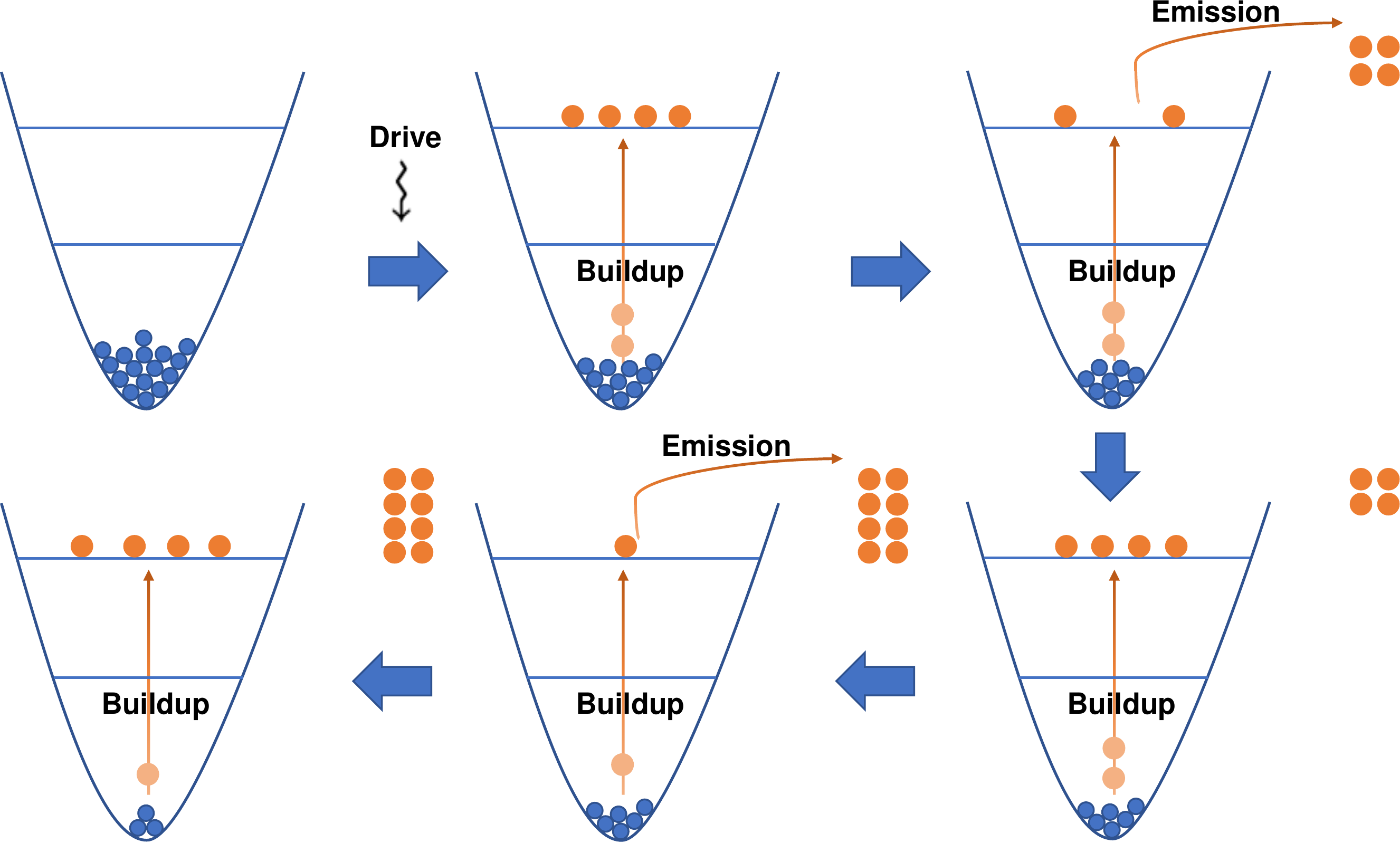}
\caption{(color online) The sketch of the emission process. The blue solid circles represent the trapped particles in the ground state, and the orange solid circles indicate the excited particles.}
\label{emiprocess}
\end{figure}

The intermittent phenomena manifested might be tentatively speculated as follows: Since there are three discrete modes, i.e., the system has three energy levels, when proper trapping potential is chosen, trapped particles are restricted in the ground state corresponding to the lowest ``P" mode at $t<0$, as sketched in Fig.~\ref{emiprocess}. Once a weak drive with related frequency is applied, particles are pumped to the dominant ``M" mode rather than directly ejected out, while building up in the corresponding energy level. Therefore, we see a short plateau in the first certain period of time, and few particles escape from the trap. To some extent, the particles reach a ``re-condensation" in this higher energy level and keep accumulation, until that the jets are instantly emitted, and an emergent ``pause" appears when the excited particles cannot maintain the dramatic emission (see the plateaus in Fig.~\ref{decaylong-insets-grid}). It is clear that the build-up and the emission are simultaneous, and instead of a real suspension of the excitation, the plateau mainly corresponds to another build-up stage. However, the emission is explicitly transient, which is much faster than the progressive build-up process. When insufficient particles are left in the higher energy level, the emission retards and the plateau shows up. With the decrease of particles in the ground state, subsequent build-up processes probably take more times, leading to a longer duration of the latter plateau. In addition, the accumulation and ejection of particles would be much quicker under larger drives, and hence multiple build-up processes and reemissions appear.

\begin{figure}[tbp]
\includegraphics[width=\columnwidth]{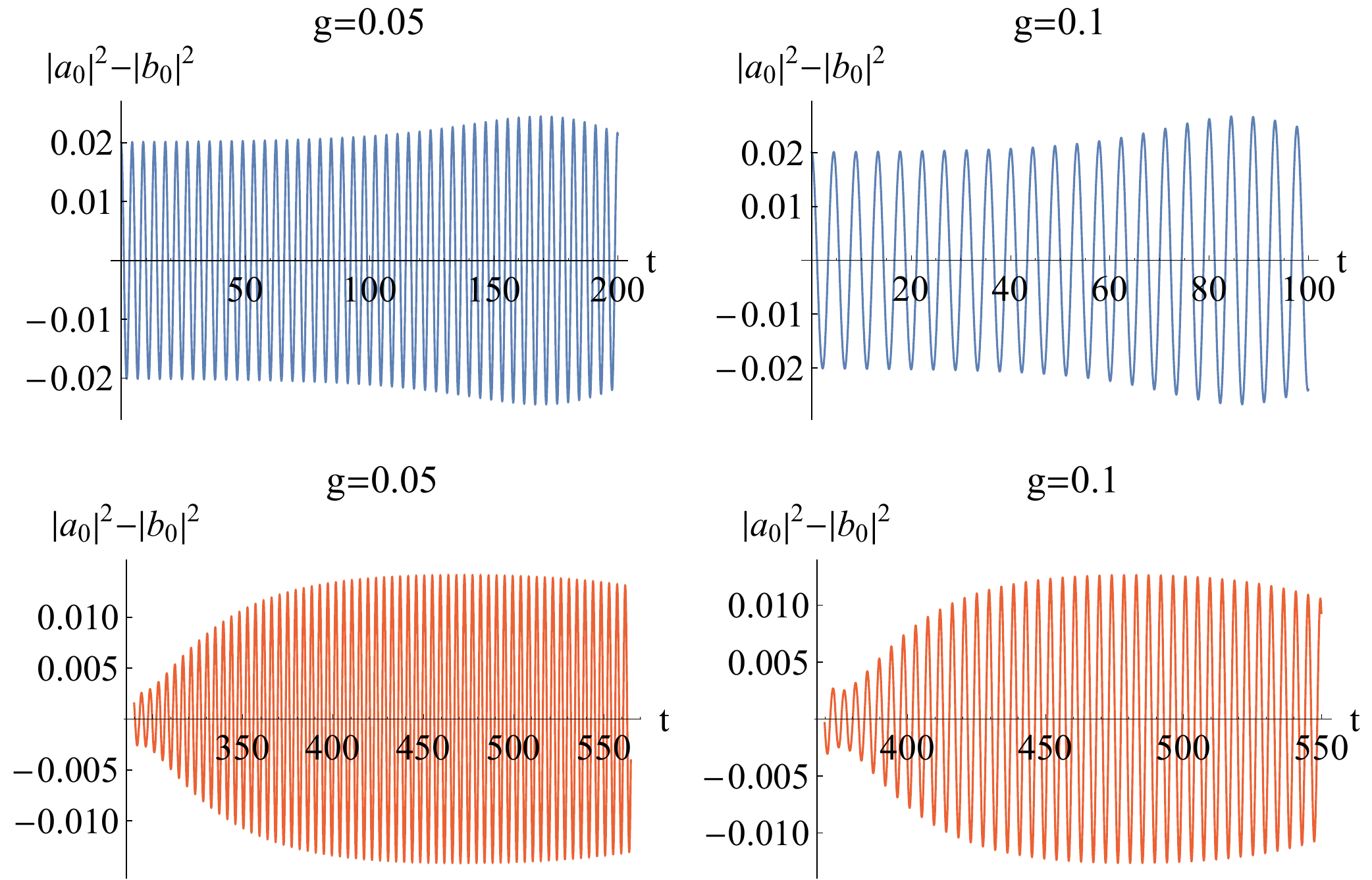}
\caption{(color online) Time evolution of the particle imbalance $|a_0|^2-|b_0|^2$ with respect to typical plateau. Upper panel: The first plateau. Lower panel: The typical plateau correspond to the insets in Fig.~\ref{decaylong-insets-grid}. Here, the trapping potential is $V=-3$, the coupling strengths are $J_h=0.1$ and $J_l=1$, and the drive frequency is $\omega=4\sqrt{2}J_c$. Energies are in units of $J_c$, and times are in units of $\hbar/J_c$.}
\label{amb}
\end{figure}

To verify the above hypothesis, we take drive strengths $g=0.05$ and $g=0.1$ as examples in Fig.~\ref{amb}, which illustrates the particle imbalance $|a_{0}|^2-|b_{0}|^2$ with respect to certain plateau. For their first plateaus, the particle imbalance of drive strength $g=0.05$ increases ($0<t<175$) before decreases with time, while the case of $g=0.1$ undergoes a shorter period ($0<t<90$), which means that the response of the system is faster under a larger drive, as expected, such that more particles are excited to the higher energy level and the conditions of emission are met earlier. As for the typical plateaus (lower panel) corresponding to the insets in Fig.~\ref{decaylong-insets-grid}, the imbalances also grow at first and then fall with time, thus within the periods particles are continuously promoted from the ground state to the higher energy level rather than a real pause, until the system restarts the emission afterwards. Within a complete emission process (e.g., g=0.1), since a number of particles have already emitted out of the trap, the latter plateau lasts for longer than the former ones, i.e., it takes a longer time for the system to be in the regime of reemission. The system, at this stage, mainly exhibits the characteristic of build-up, and hence there exist multiple plateaus.

\begin{figure}[htbp]
\includegraphics[width=\columnwidth]{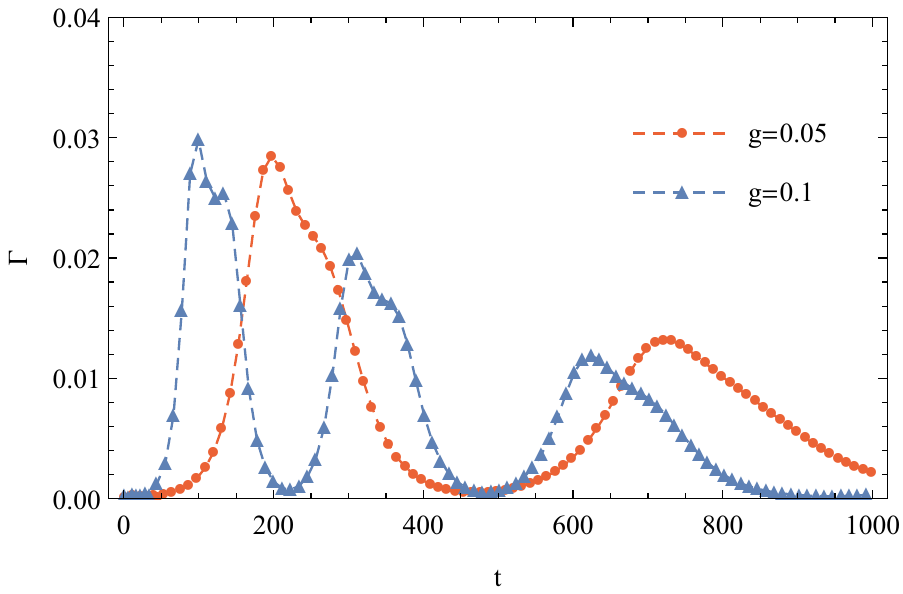}
\caption{Decay rate $\Gamma$ \textit{vs.} time $t$ for the drive strengths of $g=0.05$ and $g=0.1$, respectively. The trapping potential is $V=-3$, the coupling strengths are $J_h=0.1$ and $J_l=1$, and the drive frequency is $\omega=4\sqrt{2}J_c$. Energies are in units of $J_c$, and times are in units of $\hbar/J_c$.}
\label{decayratecom}
\end{figure}

\begin{figure}[t]
\includegraphics[width=\columnwidth]{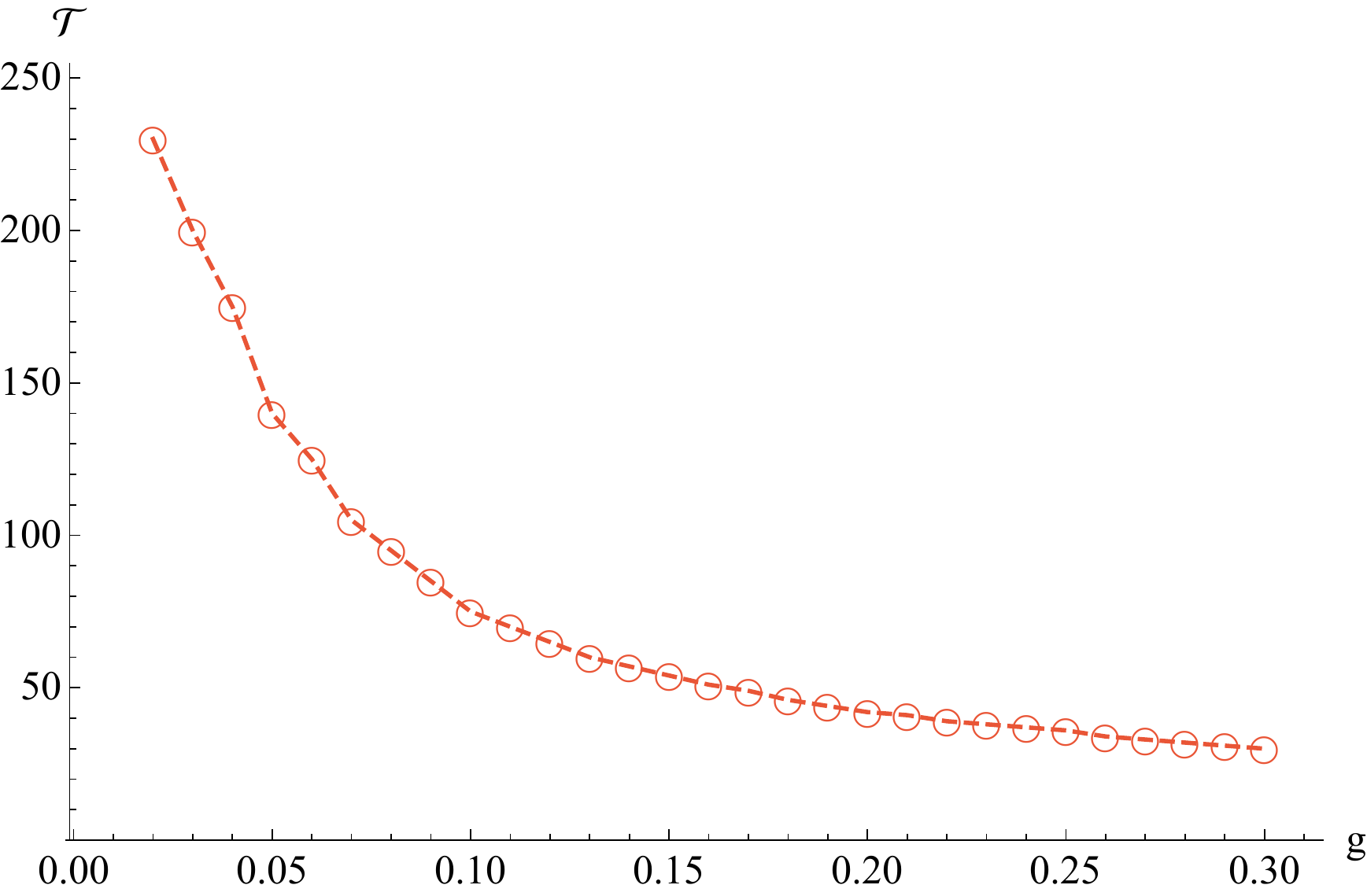}
\caption{(color online) Dependence of the duration $\mathcal{T}$ of the first plateau on the drive strength $g$. Due to the characteristic threshold, the calculations start from $g>0.02$ with drive step $\Delta g=0.01$. Here, we have taken $V=-3$, $\omega=4\sqrt{2}J_{c}$, $J_{h}=0.1$ and $J_{l}=1$. Energies are in units of $J_{c}$, and times are in units of $\hbar/J_{c}$.}
\label{duration}
\end{figure}

We also present in Fig.~\ref{decayratecom} their comparative decay rates. Both cases shortly reach their maximum before decreasing to almost zero, and then increase again to a smaller value. Within a certain period of time, the case of $g=0.1$ has a larger decay rate and ejects all the particles earlier than that of $g=0.05$, but its second maxima is somewhat less than the maximum of $g=0.05$. The lattice model is basically a multi-state system. Since we focus on a typical frequency that can pump atoms from the ground state to a high energy state, it can be simply thought of as a two-level problem here. One could, from a more insightful point of view, turn to the time-resolved survival probability of the atoms in Landau-Zener tunnelling \cite{Vitanov,Zenesini,Sinitsyn,Sinitsyn1} to better understand the stepwise structure. Although it is difficult to analytically obtain specific periodicity of the intermittency due to the nonlinearity and the staggered appearances, we are able to estimate the duration $\mathcal{T}$ of the first plateau by repeating numerical runs with tiny drive steps, as shown in Fig.~\ref{duration}, which roughly yields a decaying exponential, and reveals how long it should bear before significant jets with varying drives.

\begin{figure}[htbp]
\includegraphics[width=\columnwidth]{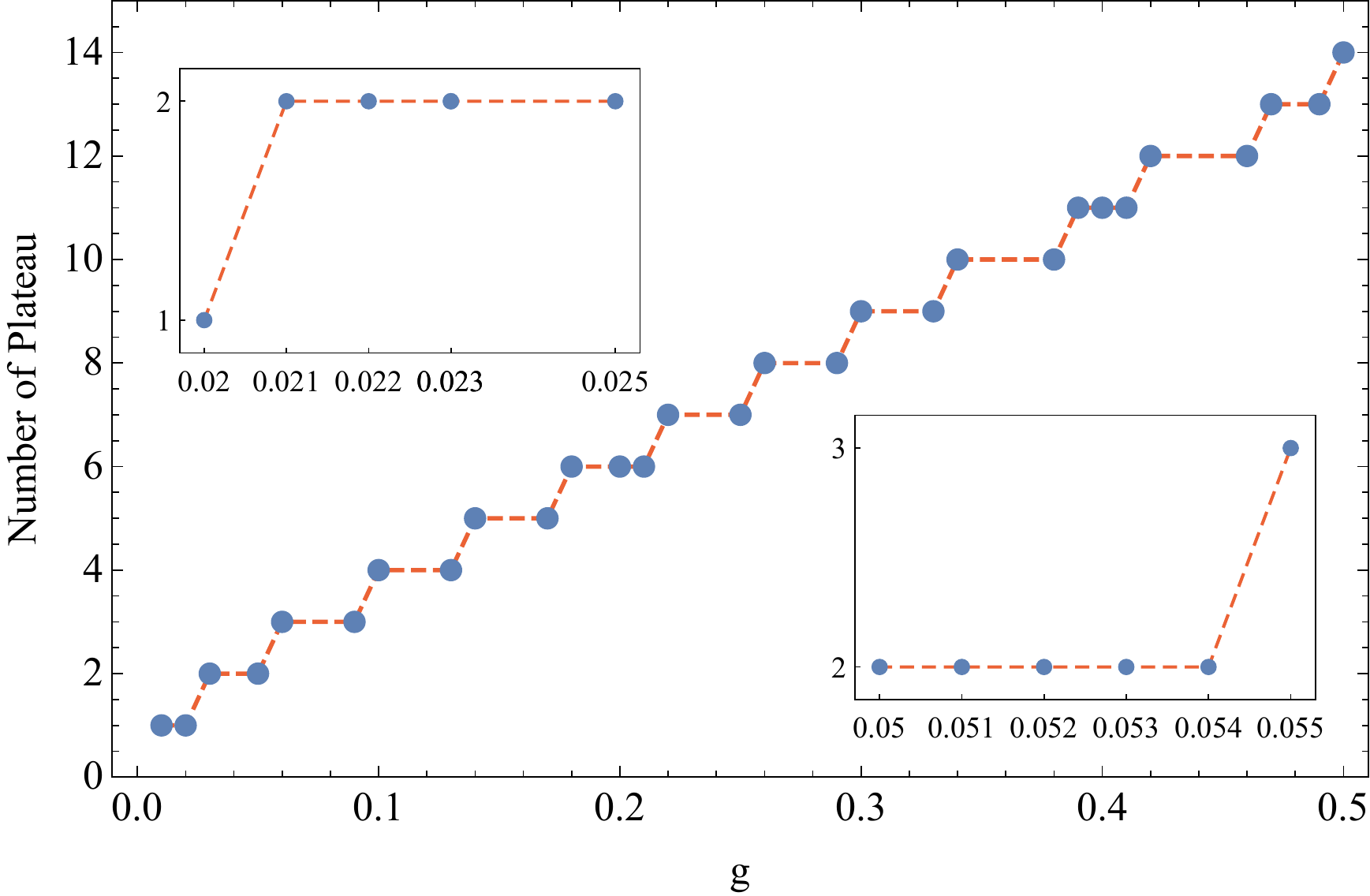}
\caption{(color online) Number of plateau as a function of the drive strength, up to time $t=1000$. Here, the trapping potential is $V=-3$, the coupling strengths are $J_h=0.1$ and $J_l=1$, and the drive frequency is $\omega=4\sqrt{2}J_c$. Energies are in units of $J_c$, and times are in units of $\hbar/J_c$.}
\label{plats}
\end{figure}

We further calculate the relation between the drive strength and the number of plateau. Since large drives may induce a series of higher-order behaviors, and make the perturbative analysis become invalid, we plainly limit ourselves to $g \leq 0.5$, and obtain an stair-like increase in Fig.~\ref{plats}. As the number of plateau cannot be non-integers, i.e., the vertical coordinate changes discontinuously while varying the drive strength continuously, we use red dashed lines to connect the numerical scatters. The two insets illustrate the details of the inflection points within $0<g<0.1$. The first jump emerges between $0.02$ and $0.021$, and the corresponding number of plateau is $1$ and $2$, respectively, while the second one appears between $0.054$ and $0.055$ with numbers of $2$ and $3$. Moreover, the number for the intervals of two inflection points should also be integers. The drive-step here is fixed as $0.001$, so we are incapable of capturing particular relations with smaller steps. The results, however, sufficiently demonstrate the dependence of the number of plateau on the drive strength, and hence illustrate the intermittency.

\begin{figure}[b]
\includegraphics[width=\columnwidth]{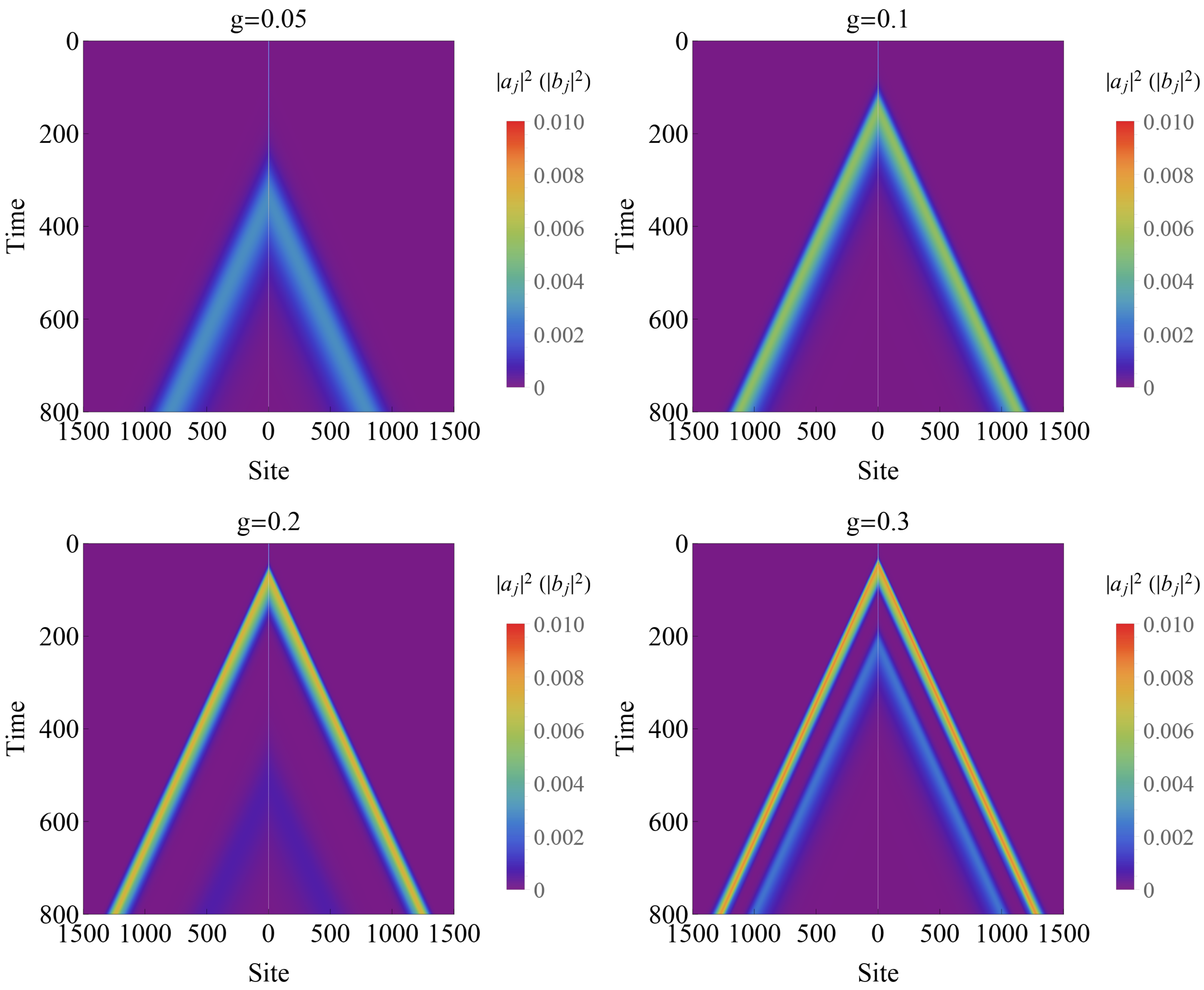}
\caption{(color online) Number of the particles on the $j$th site of each lead as a function of time, for the two-site model from Ref.~\cite{Lai2}. Here, we have taken $V=-2$, $J_c=0.1$, $J_l=1$, and $\omega=4J_{ab}$. Energies are in units of $J_{ab}$, and times are in units of $\hbar/J_{ab}$.}
\label{jetarraytot2grid}
\end{figure}

So far, by parametrically modulating the interactions we have elucidated the dynamics of intermittent emission of particles from three trapped sites constructed in Fig.~\ref{latticepic}. One can anticipate that the ``three energy level" in the current model is not necessary for observing this phenomenon.  Significant intermittent emission of particles would appear in other infinite lattices with multiple sites that share analogous geometries and couplings to this model. In the context, we immediately recall the two-site model from Ref.~\cite{Lai2}. As shown in Fig.~\ref{jetarraytot2grid}, when the drive strengths are $g=0.05$ and $g=0.1$, the resulting emissions are conventionally continuous for the drive strengths are too small to induce the intermittency. As for the drive strength of $g=0.2$, at intermediate times $100<t<200$ a large pulse of particles are emitted, with a much larger decay rate than that of $g=0.05$ and $g=0.1$, before it subsequently undergoes another build-up stage ($200<t<500$). In addition, the rate of reemission becomes fairly small since the remaining particles are quite few. The system responses much more quickly when the drive strength is further increased to $g=0.3$, and exhibits a similar intermittency to that of the former cases.

\section{Summary and Outlook}\label{sec:summary}

We have considered a one-dimensional lattice, in which the trap contains three sites, to investigate the collective emission of particles from a Bose-Einstein condensate. We parametrically modulate the interparticle interaction strength, and utilize perturbative and numerical calculations to analyze the system.

In our model, by perturbatively analyzing different modes, which leads to the conditions of being stable and stimulated, we are able to obtain two main frequencies, under which the system emits a number of particles, and find that the emission is distinctly intermittent rather than continuous. The trapped particles exhibit a stair-like decay, where a larger drive induces a more significant intermittency. The intermittency can be thought of as the build-up stage: When a weak drive with proper frequency is applied, particles in the ground sate are pumped to the corresponding higher energy level and subsequently accumulate. To some extent, the particles emit rapidly, leading to pair jets. The emission and the build-up are simultaneous, while it takes longer for the progressive build-up than that of the transient emission. When the excited particles cannot maintain the dramatic emission, the rate slows down resembling a ``pause" in the decay, but trapped particles in the ground state are still continuously promoted to the higher energy levels. Under a larger drive the regimes are satisfied earlier, inducing multiple processes of build-up and reemission.

It is interesting to note that for an infinite lattice with similar configurations and couplings, as long as there is more than one mode, with periodic modulations one could also see intermittent emissions. The framework is useful for further researches on multiple-site systems, and hence more specific investigations are in need.

\section*{Acknowledgements}
L.Q.L is grateful to Professor Erich J. Mueller for the profound instructions. This work was supported by the China Scholarship Council (Grant No.~201906130092), the Natural Science Research Start-up Foundation of Recruiting Talents of Nanjing University of Posts and Telecommunications (Grant No.~NY223065), Natural Science Foundation of Sichuan Province (Grant No.~2023NSFSC1330), and National Natural Science Foundation of China (Grant No.~11675051).

\section*{Conflict of Interest}
The authors declare no conflict of interest.

\begin{appendix}

\section{Nonzero drive strength}\label{nonzero}

Here we reintroduce the drive strength $g$ and explicitly derive the perturbative solutions in the case of a small driving field. We assume, by using the method of multiple scales, that the ansatz are
\begin{eqnarray}
a_{0}(t) &=& e^{-i\nu t}\left[\alpha(t)+gu_{a}(t)e^{-i\omega t}+gv_{a}^{*}(t)e^{i\omega t}\right],\\
b_{0}(t) &=& e^{-i\nu t}\left[\beta(t)+gu_{b}(t)e^{-i\omega t}+gv_{b}^{*}(t)e^{i\omega t}\right],\\
c_{0}(t) &=& e^{-i\nu t}\left[\gamma(t)+gu_{c}(t)e^{-i\omega t}+gv_{c}^{*}(t)e^{i\omega t}\right],
\end{eqnarray}
where $\alpha(t)$, $\beta(t)$, $\gamma(t)$, $u_{a,b,c}(t)$ and $v_{a,b,c}(t)$ are slowly varying variables, and they satisfy
\begin{eqnarray}
\int^{t}G_{11}(t-\tau)f(\tau)e^{-i\varphi\tau}d\tau \simeq f(t)e^{-i\varphi t}G_{11}(\varphi).
\end{eqnarray}

Substituting the ansatz into the intergro-differential equations (\ref{mastera})-(\ref{masterc}), and collecting terms that are linear in $g$ and proportional to $e^{\pm i\omega t}$, we reach
\begin{eqnarray}
V\alpha-J_{c}\gamma+J_{h}^{2}G_{11}(\nu)\alpha &=& \nu\alpha, \label{alpha} \\
V\beta-J_{c}\gamma+J_{h}^{2}G_{11}(\nu)\beta &=& \nu\beta, \label{beta} \\
V\gamma-J_{c}\alpha-J_{c}\beta &=&\nu\gamma, \label{gamma}
\end{eqnarray}
and
\begin{eqnarray}
\left(
\begin{array}{cccccc}
  \Delta_{+}   & 0 & 0 & 0 & J_{c} & 0 \\
     0 & \Delta_{-} & 0 & 0 & 0 & J_{c}   \\
     0  &  0 & \Delta_{+} & 0 &  J_{c} & 0   \\
     0 & 0 & 0 & \Delta_{-} & 0 & J_{c}   \\
     J_{c}  & 0 & J_{c} & 0 & \Delta_{+}^{\prime} & 0   \\
      0 & J_{c} & 0 & J_{c} & 0 & \Delta_{-}^{\prime}
\end{array}
\right)
\left(
\begin{array}{cc}
     u_{a}  \\
     v_{a}   \\
     u_{b}   \\
     v_{b}   \\
     u_{c}   \\
     v_{c}
\end{array}
\right)=\frac{i}{2}
\left(
\begin{array}{cc}
    |\alpha|^{2}\alpha   \\
    -|\alpha|^{2}\alpha^{*}  \\
    |\beta|^{2}\beta   \\
    -|\beta|^{2}\beta^{*}   \\
     |\gamma|^{2}\gamma  \\
     -|\gamma|^{2}\gamma^{*}  \\
\end{array}
\right) \nonumber \\
\end{eqnarray}
with
$\Delta_{\pm}=\nu\pm\omega-V-J_{h}^{2}G_{11}(\nu\pm\omega)$ and $\Delta_{\pm}^{\prime}=\nu\pm\omega-V$. In particular, we recover Eq.~(\ref{matrix}) through Eqs.~(\ref{alpha})-(\ref{gamma}), from which we obtain discrete modes and the regimes for inducing particle emission. The above equation can be directly inverted to get $u_{a,b,c}$ and $v_{a,b,c}$.

At second order of $g$, we start from the total particle number $N_{\rm{tot}}(t)=|a_{0}(t)|^2+|b_{0}(t)|^2+|c_{0}(t)|^2$ and $\partial_{t}|a_{0}(t)|^{2}=\alpha\partial_{t}\alpha^{*}+\alpha^{*}\partial_{t}\alpha$. As we have
\begin{eqnarray}
i\partial_{t}\alpha &=& 2|\alpha|^{2} g^{2} \left(u_{a}e^{-i\omega t}+v_{a}^{*}e^{i\omega t}\right) \frac{\left(e^{i\omega t}-e^{-i\omega t}\right)}{2i}  \nonumber \\
&&+\alpha^{2}g^{2}\left(u_{a}^{*}e^{i\omega t}+v_{a}e^{-i\omega t}\right)\frac{\left(e^{i\omega t}-e^{-i\omega t}\right)}{2i},
\end{eqnarray}
taking the ones that are winding, yields
\begin{eqnarray}
\partial_{t}|a_{0}(t)|^{2} &=& g^{2}|\alpha|^{2}
{\rm{Re}}(\alpha v_{a}-\alpha^{*}u_{a}).
\end{eqnarray}
Similar partial derivatives with respect to $|b_{0}(t)|^{2}$ and $|c_{0}(t)|^{2}$ are readily accessible.
By the neat combination of the parameters $u_{a,b,c}$ and $v_{a,b,c}$, one can calculate the dependence of the particle current on the drive strength
\begin{eqnarray}
-\partial_{t}N_{\rm{tot}} &=& g^{2}|\alpha|^{2}
{\rm{Re}}(\alpha^{*}u_{a}-\alpha v_{a}) \nonumber \\
&&+g^{2}|\beta|^{2}{\rm{Re}}(\beta^{*}u_{b}-\beta v_{b}) \nonumber \\
&&+g^{2}|\gamma|^{2}{\rm{Re}}(\gamma^{*}u_{c}-\gamma v_{c}),
\end{eqnarray}
and the corresponding average decay rate
\begin{eqnarray}
\Gamma=-\partial_{t}N_{\rm{tot}}/N_{\rm{tot}}.
\end{eqnarray}

\end{appendix}

\end{document}